\documentstyle[12pt]{article}
\begin{document}
\setcounter{page}0
\thispagestyle{empty}

\vfill

\begin{center}

{\LARGE $O(\alpha \alpha_s^2)$ correction to the electroweak
 $\rho$ parameter}

\vfill

{\large L. Avdeev\footnote{Bogoliubov Laboratory of Theoretical Physics,
 Joint Institute for Nuclear Research, 141980 Dubna (Moscow Region),
 Russian Federation.\label{LTF}}$^,$\footnote{E-mail:
 $avdeevL@thsun1.jinr.dubna.su$}$^,$\footnote{Supported by
 Volkswagen-Stiftung.\label{VW}},~ J. Fleischer\footnote{Fakult\"at
 f\"ur Physik, Universit\"at Bielefeld, D-33615 Bielefeld 1,
 Germany.\label{BU}}$^,$\footnote{E-mail:
 $Fleischer@physik.uni$-$bielefeld.de$},~ S.
 Mikhailov$^{\ref{LTF},\ref{VW},}$\footnote{E-mail:
 $mikhs@thsun1.jinr.dubna.su$},~ O.
 Tarasov$^{\ref{BU},}$\footnote{E-mail:
 $oleg@physik.uni$-$bielefeld.de$;
 {~On leave of absence from JINR, 141980 Dubna (Moscow Region), Russian
 Federation.}; {~Supported by BMFT and RFFR grant No. 93-02-14428.}}}

\end{center}

\vfill

\begin{abstract}
The three-loop QCD contributions to the vacuum polarization functions of
the $Z$ and $W$ bosons at zero momentum are calculated. The top quark is
considered to be massive and the other quarks massless. Using these
results, we calculate the correction to the electroweak $\rho$
parameter.

All computations are done in the framework of dimensional regularization
as well as regularization by dimensional reduction. We use recurrence
relations obtained by the method of integration by parts to reduce all
integrals to a small set of master integrals.

A comparison of the two-loop and three-loop QCD corrections to the $\rho$
parameter is performed. \end{abstract}

\vfill

\newpage

\setcounter{footnote}{0}

Owing to the apparent discovery of the top quark~\cite{Abe}
with a mass of $174 \pm 10$~GeV, the prospects grow to test the Standard
Model on an even higher level of precision than it was possible
by now. In particular, the precise knowledge of top-mass effects will
allow us to obtain better limits on virtual Higgs effects (and thus,
indirectly, on the Higgs mass), and possibly, on new physics. For
this reason, a great deal of work has been devoted to the study of the
top-mass effects in higher-loop radiative corrections of various
electroweak parameters. Mostly, in these studies the top mass is assumed
to be large compared to all other masses so that the latter can be put
equal to zero from the very beginning (see~\cite{VBV}). In
\cite{Barbi,FJT} also the Higgs mass $m_H$ was kept as an independent
parameter,
and both limits, $m_H\ll m_t$ and $m_H\gg m_t$, were studied.

In the Standard Model there are two different sources of corrections
which become large ($\sim G_{\mu}m_t^2$) in the limit of a heavy top,
owing to the large top-bottom mass ratio: the $Z$ and $W$ self-energies
(affecting, in particular, the $\rho$ parameter) \cite{Veltman77} and
the $Zb\overline{b}$ vertex \cite{[13]}. Experimentally, these effects
are best accessible in $e^{+}e^{-} \to f\bar{f}$ near the $Z$ resonance
measured at LEP1 and the on-resonance asymmetries measured at LEP1
and the SLAC
$e^{+}e^{-}$ linear collider SLC.

In the present paper we are concerned with the heavy-top QCD-corrections
to the electroweak $\rho$ parameter in three-loop approximation. The
$\rho$ parameter is defined as the ratio of the neutral-current to
charged-current amplitudes at zero momentum transfer:
\begin{equation}
   \rho = \frac{G_{NC}(0)}{G_{CC}(0)} = \frac{1}{1 - \Delta \rho} ,
\end{equation}
\noindent where the leading fermion contribution to $\Delta \rho$ is contained
in the gauge-boson self-energies
\begin{equation}
   \Delta \rho = \frac{\Pi_{Z}(0)}{M_Z^2} - \frac{\Pi_{W}(0)}{M_W^2} .
\label{delro}
\end{equation}
\noindent In the approximation considered, we write
\begin{equation}
  \Delta \rho =  3 x_t
 (1+\delta^{EW}+\delta^{EW,QCD}+\delta^{QCD})\simeq
  3 x_t
  (1 + \rho^{(2)} x_t)(1 + h \delta^{QCD}_{(2)}+h^2\delta^{QCD}_{(3)}) ,
\end{equation}
\noindent with
\begin{equation}
   x_t = \frac{\sqrt 2~ G_\mu }{16 {\pi}^2} m_t^2 ,~~~
  ~~~~h=\frac{\alpha_s}{4\pi},
\end{equation}
$\alpha_s$ being the QCD coupling constant. We have denoted by
$\delta^{EW}$ the pure electroweak, by $\delta^{EW,QCD}$ the mixed
electroweak-QCD,
and by $\delta^{QCD}$ the pure QCD corrections.
The two-loop electroweak correction $\rho^{(2)}$, due to virtual
Higgs (ghost) effects, is small $\rho^{(2)} |_{m_H=0} = -0.74$ for
$m_H \approx 0$  \cite{BijVe}
but reaches a maximum as large as $-$11.57 at $m_H \approx
5.7 m_t$ \cite{Barbi,FJT}.

The one-loop correction to $\Delta \rho $ was first calculated
in \cite{Veltman77}.
The two-loop QCD correction $\delta^{QCD}_{(2)}$ has been
calculated in \cite{DV}. It proved to be rather large.
If one takes $m_t$ as the top-quark pole mass, then
\begin{equation}
   \delta^{QCD}_{(2)} = -~ \frac89(\pi^2+3).
\end{equation}
\noindent Therefore, it is essential to evaluate the next, three-loop
correction, in view of the high precision of modern experiments.

To evaluate $\Delta \rho$, the diagonal parts of the self-energies of
the $W$ and $Z$ gauge bosons
\begin{equation}
\Pi^{\mu \nu}_{\alpha}(q)=g_{\mu \nu} \Pi_{\alpha}(q^2)+q_{\mu}q_{\nu}
 \tilde{\Pi}_{\alpha}(q^2)
\end{equation}
($\alpha =W,Z$) at $q=0$ are needed. Since at zero momentum and $m_t
\neq 0$ no infrared divergences appear in diagrams with only fermions
and gluons, one may put $q=0$ from the very beginning. Contracting
$\Pi^{\mu \nu}_{\alpha}(0)$ with $g_{\mu \nu}$, we obtain for
$\Pi_{\alpha}(0)$ an expression containing only bubble integrals. At the
one- and two-loop level these are quite simple and for arbitrary
space-time dimension $d=4-2\varepsilon$ can be written in terms of
Euler's $\Gamma$ function. Here we need only
\begin{eqnarray}
&&\int \frac{d^d k_1}{ \pi^{d/2}}~\frac{
 (m^2)^{\beta-d/2}}{(k_1^2+m^2)^{\beta}}
=\frac{\Gamma(\beta-d/2)}{\Gamma(\beta)},
\nonumber \\
\nonumber \\
&&\int \int \frac{d^d k_1~ d^d k_2}{ \pi^d}~\frac{
(m^2)^{\alpha+\beta+\gamma-d}}{(k_1^2+m^2)
 ^{\alpha}(k_2+m^2)^{\beta}~((k_1-k_2)^2)^{\gamma} }=
\nonumber \\
&&~~~~~~~~~
=\frac{\Gamma(\alpha+\beta+\gamma-d)~\Gamma(\frac{d}{2}-\gamma)~
\Gamma(\beta+\gamma-\frac{d}{2})~\Gamma(\alpha+\gamma-\frac{d}{2})}
{ \Gamma(\alpha)~\Gamma(\beta)~\Gamma(
\alpha+\beta+2\gamma-d)}, \nonumber\\
\nonumber\\
&&\int \int \frac{d^d k_1~ d^d k_2}{ \pi^d}\frac{
 (m^2)^{\alpha+\beta+\gamma-d}}{(k_1^2)
 ^{\alpha}((k_1-k_2)^2)^{\beta}(k_2^2+m^2)^{\gamma} }
 \nonumber \\
&&~~~~~~~~~~ =\frac{\Gamma(\alpha+\beta+\gamma-d)~
 \Gamma(\alpha+\beta-d/2)~
 \Gamma(d/2-\alpha)~ \Gamma(d/2-\beta)}{
 \Gamma(\alpha)~\Gamma(\beta)~\Gamma(\gamma)~ \Gamma(d/2)}.
\end{eqnarray}

At the three-loop level 22 diagrams of the $Z$-boson self-energy and 29
diagrams of the $W$-boson self-energy contribute to $\delta^{QCD}$.
The integrals that appear here are much more
complicated than at the one- and two-loop level.

The rather complicated task of computing massive three-loop Feynman
diagrams is accomplished by applying the method of recurrence relations
\cite{CT,David54}.
This method allows us to relate various scalar Feynman
integrals of the same prototype which differ by powers of their
scalar propagators. As a
result, by means of plain algebra, any diagram is reduced to a limited
number of so-called master integrals. They need to be evaluated once
and for all, and can then be used in any renormalizable
quantum field theory.
Some of the integrals that we need for the present three-loop calculation
were considered in \cite{David54}. Here, however,
more types of integrals are required. In addition
to the master integrals evaluated in \cite{David54},
two more
nontrivial master integrals are encountered:
\begin{eqnarray}
&\!\!\!\!\!\!\!\!& \int \int \int
\frac{d^dk_1~ d^d k_2~ d^d k_3}{[ \pi^{d/2} \Gamma(1+
  \varepsilon)]^3}~\frac{(m^2)^{4-3d/2}}{k_1^2 [(k_1-k_2)^2+m^2]
 [(k_2-k_3)^2+m^2]
 [k_3^2+m^2] }\nonumber \\*
&\!\!\!\!\!\!\!\!&~~~~~~~~  = \frac{1}{\varepsilon^3}
 +\frac{15}{4\varepsilon^2}
 +\frac{65}{8\varepsilon}+\frac{135}{16}+\frac{81}{4}S_2+O(\varepsilon),
\\
&\!\!\!\!\!\!\!\!& \nonumber \\
&\!\!\!\!\!\!\!\!& \int \int \int \frac{d^d k_1~ d^d k_2~ d^d k_3}
 {[ \pi^{d/2} \Gamma(1+\varepsilon)]^3}
 \frac{(m^2)^{6-3d/2}}{k_1^2(k_1-k_2)^2k_3^2[(k_1-k_3)^2+m^2]
 [k_2^2+m^2][(k_2-k_3)^2+m^2]} \nonumber \\*
&\!\!\!\!\!\!\!\!&~~~~~~~~ =2 \zeta(3)~ \frac{1}{\varepsilon}~+ D_3
 + O(\varepsilon)
\end{eqnarray}
with
\begin{equation}
S_2=\frac{4}{9 \sqrt{3}} {\rm Cl}_2(\frac{\pi}{3}) =
 0.260~ 434~ 137~ 632~ 162~ 098~ 955~ 729~ .
\end{equation}
We have not found a representation of $D_3$ in terms of known
transcendental numbers,
though we do not exclude its existence. By means of the numerical
method for the evaluation of Feynman diagrams proposed in \cite{FT},
$D_3$ can be calculated quite accurately. Here we give the
first 22 digits, which is more than enough for a precise evaluation of
$\rho$:
\begin{equation}
D_3=-3.027~ 009~ 493~ 987~ 652~ 019~ 786.
\end{equation}

Calculations were mostly done using FORM 1.1 \cite{FORM}. All the
diagrams were computed in the covariant gauge with an arbitrary gauge
parameter. Performing charge and mass renormalization in the
$\overline{MS}$ scheme, we got for the $W$-boson propagator the
following expression:
\begin{eqnarray}
&& \Pi_W^{(3)}(0) = 12x_t M_W^2
 \Biggl\{
 \left( -\frac{1}{2\varepsilon}-\frac14-\frac12 \hat{l} \right)
 +C_F \left(\frac{3}{2\varepsilon^2}-\frac{5}{4\varepsilon}-\frac{13}{8}
 +\zeta(2)- \hat{l}-\frac32 \hat{l}^2  \right)h \Biggr.  \nonumber \\
\nonumber  \\
&& +\Biggl[ C_F^2 \left( -\frac{3}{\varepsilon^3}
 +\frac{3}{\varepsilon^2}
 +\frac{119}{24\varepsilon}-\frac{6}{\varepsilon}\zeta(3)+\frac{1025}{72}
 +\frac{259}{18}\zeta(2)-\frac{379}{3} \zeta(3)+26\zeta(4) \right.
 \Biggr. \nonumber \\
\nonumber \\
&& \left. +\frac{1053}{4}S_2-D_3-8B_4
  +\Bigl(\frac{95}{8}+6\zeta(2)-18\zeta(3)\Bigr) \hat{l}
 +\frac{21}{4} \hat{l}^2    -3 \hat{l}^3 \right) \nonumber \\
\nonumber \\
&& +C_A C_F \left(-\frac{11}{6\varepsilon^3}
 +\frac{83}{12\varepsilon^2}
 -\frac{77}{12 \varepsilon}+\frac{3}{\varepsilon}\zeta(3)-\frac{869}{48}
 +\frac{73}{6}\zeta(2) +47\zeta(3)-21 \zeta(4)  \right. \nonumber \\
\nonumber \\
&& \left.  -\frac{1053}{8}S_2  +\frac12 D_3+4 B_4
 + \Bigl(-\frac{137}{8}+\frac{11}{3}\zeta(2)+9\zeta(3)\Bigr) \hat{l}
  -\frac{119}{12} \hat{l}^2 -\frac{11}{6} \hat{l}^3 \right) \nonumber \\
\nonumber \\
&& +C_F n_f \left(\frac{1}{3\varepsilon^3}
 -\frac{5}{6\varepsilon^2}
 +\frac{2}{3\varepsilon}+\frac{73}{24}-\frac{7}{3}\zeta(2)+4\zeta(3)
 +\Bigl(\frac94- \frac{2}{3}\zeta(2)\Bigr)\hat{l}
 +\frac{7}{6} \hat{l}^2+\frac{1}{3}\hat{l}^3 \right) \nonumber \\
\nonumber \\
&& \Biggl.\Biggl. +C_F \left(-\frac{44}{3}+\frac73\zeta(2)
 +\frac{92}{3}\zeta(3) -\frac{243}{2}S_2 \right)
 \Biggr] h^2 \Biggr\}.
\label{wms}
\end{eqnarray}
Here $n_f$ is the total number of quarks, $\hat{l}=\ln
(\mu^2/\hat{m}_t^2(\mu))$,
$\hat{m}_t(\mu)$ is the renormalized mass in the $\overline{MS}$ scheme, and
the constant
\begin{equation}
 B_4=16~ {\rm Li_4 \left(\frac12 \right)} +\frac23 \ln^4 2
 -\frac23 \pi^2 \ln^2 2 - \frac
 {13}{180} \pi^4 = - 1.762~ 800~ 087~ 073~ 770~ 086~
\end{equation}
was defined in \cite{David54}.
The result for the $Z$-boson propagator is
\begin{eqnarray}
&& \Pi_Z^{(3)}(0) = 12x_tM_Z^2
 \Biggl\{
 \left( -\frac{1}{2\varepsilon}-\frac12 \hat{l} \right)
+C_F \left( \frac{3}{2\varepsilon^2}-\frac{5}{4\varepsilon}-\frac18
 +\frac 1 2 \hat{l} -\frac 3 2 \hat{l}^2 \right) h
 \Biggr. \nonumber \\
&&\nonumber \\
&& +\Biggl[ C_F^2 \left( -\frac{3}{\varepsilon^3}+\frac{3}{\varepsilon^2}
 +\frac{119}{24\varepsilon}-\frac{6}{\varepsilon}\zeta(3)+\frac{51}{16}
 -36 \zeta(3)+27\zeta(4)-6B_4 \right. \Biggr. \nonumber \\
&& \nonumber \\
&& \left. + \Bigl( \frac{101}{8}-18\zeta(3) \Bigr) \hat{l}
 +\frac{39}{4}\hat{l}^2 -3 \hat{l}^3 \right) \nonumber \\
&& \nonumber \\
&&+C_A C_F \left(-\frac{11}{6\varepsilon^3}+\frac{83}{12\varepsilon^2}
 -\frac{77}{12\varepsilon}+\frac{3}{\varepsilon}\zeta(3)+3+\frac{28}{3}
\zeta(3)-\frac{27}{2}\zeta(4)+3B_4 \right. \nonumber\\
&&\nonumber \\
&&+\left. \Bigl(-\frac{85}{24}+9\zeta(3)\Bigr) \hat{l}
  -\frac{43}{6} \hat{l}^2 -\frac{11}{6} \hat{l}^3  \right) \nonumber \\
&& \nonumber \\
&&+C_F n_f \left(\frac{1}{3\varepsilon^3}-\frac{5}{6\varepsilon^2}
 +\frac{2}{3\varepsilon}-\frac{1}{12}+\frac{8}{3}\zeta(3)
 + \frac{5}{12}\hat{l}+\frac{2}{3} \hat{l}^2
 +\frac{1}{3}\hat{l}^3 \right)
 \nonumber \\
&& \nonumber \\
&& \Biggl. \Biggl. +C_F \Bigl( -2-12\zeta(3) \Bigr)  \Biggr] h^2 \Biggr\}.
\label{zms}
\end{eqnarray}

In the sum of the bare diagrams contributing to $\Pi_{\alpha}(0)$
($\alpha=W,Z$),
as well as in the corresponding counterterms separately, the
gauge parameter cancels,
which is a partial check of our result.

Substitution of (\ref{wms}) and (\ref{zms}) into (\ref{delro}) gives
the ultraviolet-finite expression

\begin{eqnarray}
\delta^{QCD}_{(3),\overline{MS}}&=&
C_F^2\Biggl[ -\frac{1591}{36}-\frac{518}{9}\zeta(2)
 +\frac{1084}{3}\zeta(3)+4\zeta(4)
 -1053S_2 +4D_3 +8B_4 \Biggr.\nonumber \\
&& \nonumber\\
&& \Biggl. +\Bigl(3-24 \zeta(2)\Bigr) \hat{l} +18 \hat{l}^2 \Biggr]
 \nonumber \\
&& \nonumber \\
&& +C_A C_F \left[ \frac{1013}{12} -\frac{146}{3}\zeta(2)
 -\frac{452}{3}\zeta(3)
 +30\zeta(4)+\frac{1053}{2}S_2 \right. \nonumber \\
&& \nonumber \\
&& \left. -2D_3-4B_4+
 \Bigl(\frac{163}{3}-\frac{44}{3}\zeta(2)\Bigr) \hat{l}
 +11 \hat{l}^2 \right] \nonumber \\
&& \nonumber \\
&& +C_F n_f \left[-\frac{25}{2}+\frac{28}{3}\zeta(2)-\frac{16}{3}\zeta(3)
 +\Bigl(-\frac{22}{3}+\frac83\zeta(2)\Bigr) \hat{l} -2 \hat{l}^2 \right]
 \nonumber \\
&& \nonumber \\
&& +C_F \left[\frac{152}{3}-\frac{28}{3}\zeta(2)-\frac{512}{3}
 \zeta(3)+486 S_2 \right].
\label{del3ms}
\end{eqnarray}

If we perform mass renormalization in such a way that the renormalized
mass is the pole mass $m_t$~, then we obtain
\begin{eqnarray}
\label{del3}
\delta^{QCD}_{(3)}&=&
 C_F^2 \Biggl[ -\frac{238}{9}-\frac{770}{9}\zeta(2)
 +96\zeta(2)\ln 2 +\frac{1012}{3}\zeta(3)+4\zeta(4)
 \Biggr. \nonumber \\
 \nonumber\\
&& \Biggl. -1053S_2 +4D_3 +8B_4 \Biggr] \nonumber \\
 \nonumber \\
&& +C_A C_F \left[ -\frac{49}{6}-\frac{98}{3}\zeta(2)
 -48\zeta(2) \ln 2 -\frac{416}{3}\zeta(3)
 +30\zeta(4)+\frac{1053}{2}S_2 \right. \nonumber \\
 \nonumber \\
&& \left. -2D_3-4B_4+ \Bigl(-\frac{22}{3}-\frac{44}{3}\zeta(2)\Bigr) l
 \right] \nonumber \\
 \nonumber \\
&&+C_F n_f \left[-\frac23+\frac{52}{3}\zeta(2)-\frac{16}{3}\zeta(3)
 +\Bigl(\frac43+\frac83\zeta(2)\Bigr) l \right] \nonumber \\
 \nonumber \\
&&+C_F \left[\frac{188}{3}-\frac{100}{3}\zeta(2)-\frac{512}{3}
 \zeta(3)+486 S_2 \right] .
 \end{eqnarray}
In the above formula $l=\ln(\mu^2 /m_t^2)$ .
Ultraviolet finiteness of (\ref{del3ms}) and (\ref{del3})
is an additional check of our result.
Expression (\ref{del3}) can also be obtained from our result in
the $\overline{MS}$ scheme
by using the relation between $\hat{m}_t$ in the $\overline{MS}$
scheme and the pole mass $m_t$ \cite{Gray}.

In the $\overline{MS}$ scheme for QCD [i.e. for the SU(3) gauge
group with $C_A=3$, $C_F=4/3$] we get the following expression:
\begin{eqnarray}
&&\delta^{QCD}_{\overline{MS}}=\left( 8-\frac{16}{3}
\zeta(2)+8 \hat{l} \right) h
+\left[ \frac{26459}{81}-\frac{16}{9}B_4-\frac{25064}{81} \zeta(2)
 -\frac{5072}{27}\zeta(3)
\right. \nonumber \\
&& \nonumber \\
&& +\frac{1144}{9}\zeta(4)
+882 S_2-\frac{8}{9} D_3
+n_f\left( -\frac{50}{3}+\frac{112}{9} \zeta(2) -\frac{64}{9}
 \zeta(3) \right) \nonumber \\
&& \nonumber \\
&& \left.+ \left( \frac{668}{3}-\frac{304}{3} \zeta(2)
 +\Bigl(-\frac{88}{9}+\frac{32}{9}
 \zeta(2) \Bigr) n_f \right) \hat{l}
 +\left( 76-\frac{8}{3}n_f \right) \hat{l}^2 \right] h^2.
\label{massms}
\end{eqnarray}

Substituting numerical values for all the constants
and taking $\mu^2=\hat{m}^2_t$ with the minimally subtracted mass
we obtain
\begin{equation}
\delta^{QCD}_{\overline{MS}}=-0.061\,511\,928\,430\,2~ \alpha_s
 -(0.221\,937\,307\,314\,4 +0.030\,043\,860\,323\,8~n_f)~ \alpha^2_s ,
\end{equation}
which at $n_f=6$ turns into
\begin{equation}
\delta^{QCD}_{\overline{MS}}=-0.061\,511\,928\,430\,2~ \alpha_s
 +0.402\,200\,469\,257\,5~ \alpha_s^2 .
\end{equation}
The smallness of this correction in the $\overline{MS}$ scheme
confirms expectations about higher order effects in electroweak
parameters (see e.g. \cite{FJ}).

With the definition of the renormalized mass $m_t$ as the pole mass
of the top quark we obtain:
\begin{eqnarray}
\delta^{QCD}&=&-\frac23\Bigl(1+2\zeta(2)\Bigr) \frac{\alpha_s} {\pi}
 + \left[ \frac{157}{648}
 -\frac{3313}{162}\zeta(2)-\frac{308}{27}\zeta(3)  \right.
 \nonumber \\
 \nonumber\\
&&+\frac{143}{18}\zeta(4) -\frac43 \zeta(2) \ln 2
 +\frac{441}{8}S_2-\frac19  B_4
 -\frac{1}{18}D_3   \nonumber \\
 \nonumber\\
&& \left.
 -\left(\frac{1}{18}-\frac{13}{9}\zeta(2)+\frac49 \zeta(3)\right) n_f
 -\left( \frac{11}{6}-\frac{1}{9}n_f \right)
 \Bigl(1+2\zeta(2)\Bigr) l \right]  \left(\frac{ \alpha_s}{\pi}\right)^2.
\label{mpole}
\end{eqnarray}

Substituting numerical values for all the constants and putting
$\mu^2=m^2_t$ we get
\begin{equation}
 \delta^{QCD}=-0.910\,338\,291\,586\,9~ \alpha_s
 -(2.564\,571\,412\,664\,2 -0.180\,981\,195\,767\,9~n_f)~ \alpha^2_s ,
\end{equation}
and at $n_f=6$ we have
\begin{equation}
 \delta^{QCD}=-0.910\,338\,291\,586\,9~ \alpha_s
 -1.478\,684\,237\,779\,7~ \alpha^2_s.
\label{delnumbers}
\end{equation}

Both two- and three-loop QCD contributions are negative, and their
effect is a screening of the bare mass splitting, so that only a reduced
``effective'' quantity enters the $\rho$ parameter.

Following the method of fastest apparent convergence \cite{FAC}, we can
absorb our three-loop correction into a rescaling of $\alpha_s$.  With
$n_f=6$, $\delta^{QCD}_{(3)}$ will be zero, if we take $\mu \simeq
0.2327~m_t$.  When we apply the BLM procedure \cite{BLM}, the
$n_f$-dependent term in (\ref{mpole}) can be absorbed into the
rescaling of $\alpha_s$, if we choose $\mu \simeq 0.154~m_t$.  The same
value was also obtained in \cite{SM}. We conclude that the expression
for the term proportional to $n_f$, given in \cite{SM}, agrees with the
analytical result for this term given in (\ref{mpole}).  As an
important result of our calculation, we stress the stability of
$\delta^{QCD}$: the usual perturbation theory, the FAC and BLM
procedures give rather close results for $\delta^{QCD}$.  Taking
$\alpha_s(m_t)=0.1055$, we have $\delta^{QCD}=-$0.1125, $-$0.1159 and
$-$0.1154 for the perturbation theory, FAC and BLM procedures,
respectively.

For our calculations we used the anticommuting $\gamma_5$. To check our
result obtained with this prescription at least partially, we calculated
$\delta^{QCD}$ again, using the regularization by dimensional reduction
\cite{RDR} which keeps the algebra of $\gamma$ matrices
four-dimensional. For propagator-type diagrams in the three-loop
approximation the inconsistency of this recipe is not yet revealed. As
was expected, the result that we obtained in the regularization by
dimensional reduction agrees with $\delta^{QCD}$ in the conventional
dimensional regularization after the following recalculation of the
coupling constant:
\begin{equation}
h_{RDR}=h \Bigl( 1+\frac13 C_A h \Bigr),
\end{equation}
which just corresponds to a change in the renormalization scheme. The
relation can be derived, for example, by equating invariant charges in
these two schemes.

Special attention was paid to the evaluation of the diagram with the
axial anomaly. Only one such diagram contributes to $\Delta \rho$
(namely, to $\Pi_Z(0)$) with two triangles and only the top quark
running around. We evaluated it in the framework of the regularization
by dimensional reduction and using the prescription given in
\cite{Larin}. The contribution from this diagram is finite, and the
calculations yielded the same value in both approaches (although the
evanescent parts were different). Our result also agrees with the one
given in \cite{anomaly}. At $n_f=6$ this contribution amounts to about
30\% of the total three-loop correction (\ref{delnumbers}).

Some observables that are affected by our result are briefly mentioned
here. One of them is the mass of the $W$ boson as predicted from
$\alpha$, $G_\mu$ and $M_Z$ \cite{CHJ}
\begin{equation}
M_W^2 =  \frac{\rho M_Z^2}{2}\;
\left(1+\sqrt{1-\frac{4A^2_0}{\rho M_Z^2}
\Bigl(\frac{1}{1-\Delta \alpha}+\ldots\Bigr)}\; \right),
\end{equation}
where
$
A_0=\left( \frac {\displaystyle \pi \alpha} {\sqrt 2 G_\mu} \right) ^{1/2}
=37.2802(3)\;
$
and $\Delta \alpha \simeq 0.06$ is the shift of the fine structure constant
$\alpha$ due to photon vacuum polarization effects. The ellipsis stands for the
non-leading remainder terms.
Another physical quantity is the effective weak mixing parameter relevant
to $Z$-resonance physics. It is given by
\begin{equation}
\sin^2 \bar{\Theta} =1-\frac{M_W^2}{\rho M_Z^2} = \frac{1}{2}
\left( 1-
 \sqrt{1-\frac{4A_0^2}{\rho M_Z^2}
  \Bigl(\frac{1}{1-\Delta\alpha}+\ldots\Bigr)}\;
\right) .
\end{equation}
Numerical values illustrating how these observables are affected by
various corrections are given in Table~1. QCD corrections were
calculated using (\ref{delnumbers}).

\begin{table}[htbp]
\caption{Percentage of various two- and three-loop QCD heavy-top
corrections at $\alpha_s(m_t)=0.1055$ and $m_t=174$~GeV. The value of
$\alpha_s$ was obtained by extrapolating
$\alpha_s(M_Z$=91.1895~GeV)=0.118 to the scale $\mu=m_t$ with the aid of
the three-loop $\beta$ function \protect\cite{Tar} with $n_f=5$.}

\begin{center}
\begin{tabular}{|c|r|r|rrrr|}
\hline
Observable & Two-loop  & Three-loop  & \multicolumn{4}{|c|}
 {Two-loop electroweak}\\
&\multicolumn{1}{|c|}{QCD} & \multicolumn{1}{|c|}{QCD} &
 $m_H/m_t=0$ & 1.5 & 5.7 & 10 \\
\hline
$M_W $& $-$0.065~~~ & $-$0.011~~~~ & $-$0.002 & $-$0.018 & $-$0.025
 & $-$0.023\\
$\sin^2 \bar{\Theta} $& 0.130~~~ &  0.022~~~~ &  0.003 &  0.036 &
 0.051 &  0.046\\
\hline
\end{tabular}
\end{center}
\end{table}

 Table~1 demonstrates that the three-loop QCD correction is comparable
with the two-loop electroweak correction for sizable Higgs masses (for
$m_H/m_t$=1.5 the former amounts to more than 60\% of the latter).
Thus, we conclude that it makes sense to evaluate subleading
electroweak two-loop corrections to $\Delta \rho$ (or another
observable) only if the three-loop QCD corrections are taken into
account as well.

\bigskip
{\em Acknowledgments. }
L.Avdeev, S.Mikhailov and O.Tarasov are grateful to the Physics
Department of the Bielefeld University for warm hospitality. L.Avdeev
and S.Mikhailov are thankful to the Volkswagen-Stiftung and O.Tarasov to
the BMFT and RFFR grant for financial support.
The authors are grateful to F.Jegerlehner for carefully reading the
manuscript.


\end{document}